\documentclass[11pt,a4paper]{article}
\usepackage{jcappub}
\usepackage{amsmath,amssymb,graphics,epsfig,subfigure}
\usepackage{color}

\title{Shadow of noncommutative geometry inspired black hole}

\author{Shao-Wen Wei,}
\author{Peng Cheng,}
\author{Yi Zhong,}
\author{Xiang-Nan Zhou}
\affiliation{Institution,\\Theoretical Physics, Lanzhou University, Lanzhou 730000, People's Republic of China}

\emailAdd{weishw@lzu.edu.cn, pcheng14@lzu.edu.cn, zhongy13@lzu.edu.cn, zhouxn10@lzu.edu.cn}

\abstract{
In this paper, the shadow casted by the rotating black hole inspired by noncommutative geometry is investigated. In addition to the dimensionless spin parameter $a/M_{0}$ with $M_{0}$ black hole mass and inclination angle $i$, the dimensionless noncommutative parameter $\sqrt{\vartheta}/M_{0}$ is also found to affect the shape of the black hole shadow. The result shows that the size of the shadow slightly decreases with the parameter $\sqrt{\vartheta}/M_{0}$, while the distortion increases with it. Compared to the Kerr black hole, the parameter $\sqrt{\vartheta}/M_{0}$ increases the deformation of the shadow. This may offer a way to distinguish noncommutative geometry inspired black hole from Kerr one via astronomical instruments in the near future.}


\begin{document}


\maketitle

\section{Introduction}
\label{secIntroduction}

Due to the quantum effects near the event horizon, a general relativity (GR) black hole was found to possess Hawking radiation \cite{Hawking}, which is very different from the classical concept that nothing can escape from a black hole. Subsequently, four laws of black hole thermodynamics were established. They greatly improved our understanding of the black hole. However, singularity of spacetime is one of the challenging problems in GR.

Many attempts were proposed to give a singularity free theory. For example, introduce the higher derivative covariant generalizations of GR \cite{Banados,Biswas}, which is required to be well behaved in the ultraviolet and reduced suitably to Einstein's gravity in the infrared. Other methods suggest introducing a minimal length to cure the singularity in a small distance, which means that the singularity or ultraviolet divergence is a spurious effect resulting from the inadequacy of the formalism at small distance scales \cite{Calmet,Garattini}. Noncommutative geometry is shown to be a good candidate to implement a minimal length in physical theories.

From the fundamental view of noncommutative geometry, the picture of spacetime as a manifold of points breaks down at Planck length. And a particle cannot be accurately localized in the quantum phase space \cite{Akofor}, since $[\hat{x}_{i},\hat{p}_{j}]=i\hbar\delta_{ij}$. Therefore, it is rational to replace a point in classical commutative manifold with a state in a noncommutative algebra. And a point-like object will be replaced with a smeared one \cite{Smailagic}. In general, there are two ways to construct the noncommutative quantum field theory: one is based on the Weyl-Wigner-Moyal $*$-product and the other on the coordinate coherent state formalism \cite{Smailagic}. Following the latter way, one has $[\hat{x}^{\mu}, \hat{x}^{\nu}]=i\vartheta \text{diag}(\epsilon_{1}, {\cdots}, \epsilon_{D/2})$. Then the Lorentz invariance can be recovered by imposing a single noncommutative parameter $\vartheta$, and unitarity was also checked at the one-loop level with no violation \cite{Smailagic2}.

Inspired by noncommutative geometry, the key idea to construct singularity-free black holes is replacing the point-like mass located at the center singularity with a smeared mass \cite{Smailagic}. The scale of the distribution of the mass density is denoted with $\sqrt{\vartheta}$, which has a dimension of length. That parameter is a new fundamental length scale on the same ground as Planck length $l_{P}$. In order to keep some degree of generality, we consider it as a free, model-dependent parameter, and one can call it noncommutative parameter. Motivated by the result, various black hole solutions inspired by noncommutative geometry were found \cite{Nicolini,Nicolini2,Nicolini3,Nicolini4}, and these black hole solutions were extensively studied. At the terminal stage of the black hole evaporation, the singularity problem can also be well cured \cite{Nicolini}. Other properties such as thermodynamic properties and black hole lensing were examined (see Refs. \cite{Nozari,Ding,DingLiu} and references therein). All the results show noncommutative parameter dependent effects. And these black holes become the richest class of quantum gravity black holes \cite{Mann}. In Ref. \cite{Ding}, the authors showed that the noncommutative parameter is measurable when $\sqrt{\vartheta}/M_{0}\in(0.24, 0.52)$. Therefore, testing the existence of the noncommutative parameter through black hole lensing will shed further light on such theory. However, the image of the source is a one-dimensional light ring or arc making the test of $\sqrt{\vartheta}/M_{0}$ more difficult.

Compared with it, the black hole shadow is a two-dimensional dark zone seen from the observer. So it can be easily observed from astronomical observation. By future interferometers \cite{Cash,Hirabayashi,Doeleman,Grenzebach}, direct observation of a black hole will become possible in the near future. Black hole shadow is the optical appearance casted by a black hole with the photon sources located at infinity and distributed uniformly in all directions. For a nonrotating black hole, the shadow appears as a perfect circle. While for a rotating black hole, it will be elongated in the direction of the rotation axis due to the dragging effect \cite{Chandrasekhar}. The size and distortion of the shadow are found to depend on the parameters of the black hole and the location of the observer. So the test of the black hole shadow provides an efficient method to test the nature of a black hole in the near future. And the topic has been examined by several groups in the last few years \cite{Falcke,Takahashi,Hioki,Bambi,Kraniotis,Bozzagrg,Schee,Maeda,Amarilla,Stuchlik,AmarillaEiroa,
YumotoNitta,Amarilla13,Nedkova,Wei,Tsukamoto,Atamurotov,Atamurotov2}.

So the main purpose of this paper is to investigate the effect of the dimensionless noncommutative parameter $\sqrt{\vartheta}/M_{0}$ on the black hole shadow. Using the null geodesic, we obtain the observables proposed in Refs. \cite{Maeda,Tsukamoto}. The shape of the shadow is found to closely depend on $\sqrt{\vartheta}/M_{0}$. And the result suggests that, compared with the Kerr black hole, the test of $\sqrt{\vartheta}/M_{0}\in(0.4,0.52)$ through the observation of the black hole shadow is possible in the near future using the very-long baseline interferometry (VLBI).

The paper is structured as follows. In Sec. \ref{geodesics}, we give a brief review of the null geodesics for the noncommutative geometry inspired black holes. With the effective potential, the unstable circular photon orbits are studied in Sec. \ref{potential}. Then in Sec. \ref{shape}, we get the apparent shape of the shadow casted by the black hole with or without the spin. The observables are also obtained, and the comparison is made between the noncommutative geometry inspired black hole and the Kerr black hole. The summary and remarks follow in Sec. \ref{Discussion}. We adopt the geometric units, i.e., $\hbar=c=k_{B}=G=1$.

\section{Null geodesics of noncommutative geometry inspired black hole}
\label{geodesics}

The key point to construct a noncommutative geometry inspired black hole is replacing the point-like object with a smeared one. Therefore, in a spherically symmetric case, the mass density of such an object is modified as a Gaussian density
\begin{eqnarray}
 \rho_{G}=\frac{M_{0}}{(4\pi\vartheta)^{3/2}}e^{-r^{2}/4\vartheta},
\end{eqnarray}
where $M_{0}$ is the total mass of the object and $\vartheta$ measures the distribution of the mass. In the spheroidal coordinates of the Minkowski spacetime, the mass density for a rotating object is \cite{Gurses,Dymnikova}
\begin{eqnarray}
 \rho_{M}=\frac{r^{2}}{r^{2}+a^{2}\cos^{2}\theta}\rho_{G}.
\end{eqnarray}
Then the mass encoded in a radius $r$ will be
\begin{eqnarray}
 { M(r)}&=&2\pi\int_{0}^{r}dr\int_{0}^{\pi}d\theta\sin\theta (r^{2}+a^{2}\cos^{2}\theta)\rho_{M}\nonumber\\
 &=&\frac{2M_{0}}{\sqrt{\pi}}\gamma(3/2, r^{2}/4\vartheta),\label{mass}
\end{eqnarray}
where $\gamma(3/2, x)=\int_{0}^{x}t^{1/2}e^{-t}dt$. Thus, the energy-momentum tensor for the generalized Kerr black hole reads \cite{Nicolini4}
\begin{eqnarray}
 T^{\mu}_{\;\nu}=(\rho+p_{\theta})(u^{\mu}u_{\nu}-l^{\mu}l_{\nu})-p_{\theta}\delta^{\mu}_{\;\nu},
\end{eqnarray}
where $u^{\mu}=\sqrt{-g^{rr}}(\delta^{\mu}_{\;t}+\frac{a}{(a^{2}+r^{2})}\delta^{\mu}_{\;\phi}) $
and $l^{\mu}=-\frac{1}{\sqrt{-g_{rr}}}\delta^{\mu}_{\;r}$. The invariant energy density $\rho=r^{4}\rho_{G}/(r^{2}+a^{2}\cos^{2}\theta)^{2}$. Imposing $\nabla_{\mu}T^{\mu\nu}=0$, one can obtain the pressure $p_{\theta}$.

\begin{figure*}
\begin{center}
\subfigure[]{\label{pmetric}
\includegraphics[width=8cm,height=6cm]{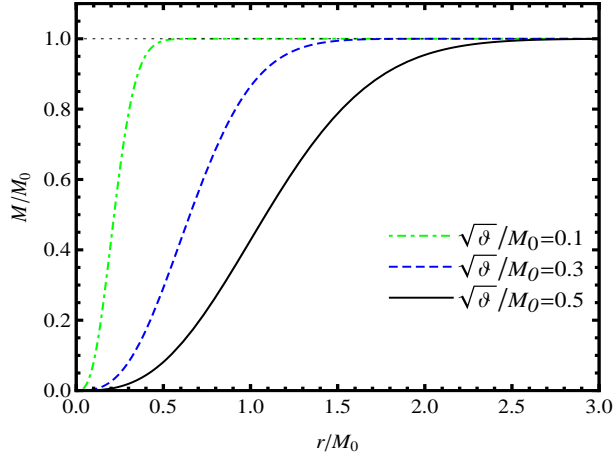}}
\subfigure[]{\label{phorizon}
\includegraphics[width=8cm,height=6cm]{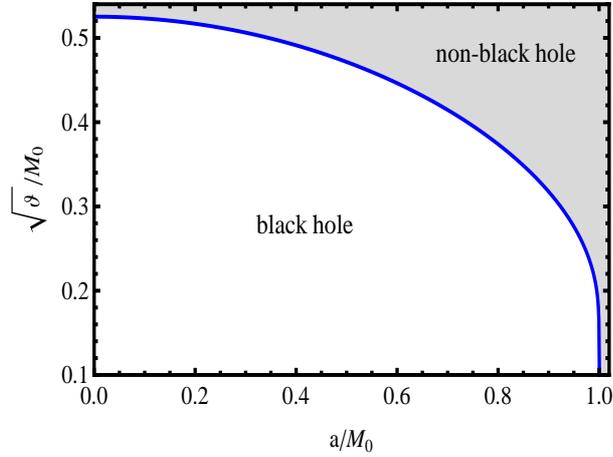}}
\end{center}
\caption{(a) Mass distribution for different values of the noncommutative parameter $\sqrt{\vartheta}/M_{0}$. (b) The parameter spaces of $\sqrt{\vartheta}/M_{0}$ and $a/M_{0}$ describing a black hole and a non-black hole. The boundary described by the blue line denotes the extremal black holes.}
\end{figure*}

Finally, taking the choice of a ``noncommutative geometry inspired" matter source as the input of the Einstein equations, Smailagic and Spallucci obtained a metric of the noncommutative geometry inspired Kerr black hole \cite{Nicolini4}:
\begin{eqnarray}
 ds^{2}=&-&\frac{\Delta-a^{2}\sin^{2}\theta}{\rho^{2}}dt^{2}
        -\frac{2a(r^{2}+a^{2}-\Delta)\sin^{2}\theta}{\rho^{2}}dtd\phi\nonumber\\
       &+&\frac{\rho^{2}}{\Delta}dr^{2}+\rho^{2}d\theta^{2}
        +\frac{(r^{2}+a^{2})-\Delta a^{2}\sin^{2}\theta}{\rho^{2}\csc^{2}\theta} d\phi^{2},~~~\label{metric}
\end{eqnarray}
where the metric functions are given by
\begin{eqnarray}
 \rho^{2}=r^{2}+a^{2}\cos^{2}\theta, \;\; \Delta=r^{2}-2Mr+a^{2},
\end{eqnarray}
$a$ is the spin parameter of the spacetime, and $M$ is given in (\ref{mass}) and denotes the mass encoded in a sphere with radius $r$. In the limit $\sqrt{\vartheta}/M_{0}\rightarrow 0$, this metric describes the conventional Kerr spacetime. In Fig. \ref{pmetric}, we show the mass distribution function $M(r)$ as a function of $r$. One can see that the mass dominates in small value of $r$ for a small value of $\sqrt{\vartheta}/M_{0}$. It is clear that $\sqrt{\vartheta}/M_{0}$ measures the noncommutativity of spacetime and it is reasonable to treat it as the noncommutative parameter of spacetime as suggested in Ref. \cite{Ding}. It is also worth to note that, for different values of $\sqrt{\vartheta}/M_{0}$ and $a$, the metric displays different horizon structures, i.e., two horizons, one horizon, and no horizon. In this paper, we only focus on the case that the black hole has, at least, one horizon, thus for fixed value of the noncommutative parameter $\sqrt{\vartheta}/M_{0}$, the spin parameter is required to $0\leq a\leq a_{max}$, where $a_{max}$ can be obtained by numerically solving $\Delta=0$ and $\partial_{r}\Delta=0$. On the other hand, given a value of spin $a/M_{0}$ for a black hole, there will be a maximum $\sqrt{\vartheta}/M_{0}$. We show the parameter space that corresponds to black holes in Fig. \ref{phorizon}. Thus, for a black hole, the parameters must be in the range
\begin{eqnarray}
 \sqrt{\vartheta}/M_{0}\in(0,0.52),\quad a/M_{0}\in(0,1).\label{range}
\end{eqnarray}
We give a fitting form of the parameters for the extremal black hole (thick blue line in Fig. \ref{phorizon}),
\begin{eqnarray}
 \sqrt{\vartheta}/M_{0}&=&0.52+0.23a/M_{0}-6.91 (a/M_{0})^2+84.52 (a/M_{0})^3
    -560.97 (a/M_{0})^4+2177.02 (a/M_{0})^5\nonumber\\
    &-&5193.85 (a/M_{0})^6 +7711.70 (a/M_{0})^7-6941.67 (a/M_{0})^8+3465.86 (a/M_{0})^9-736.26 (a/M_{0})^{10},\label{thth}
\end{eqnarray}
Then the maximum $a/M_{0}$ or $\sqrt{\vartheta}/M_{0}$ can be obtained with this relation. In the rest of the paper, we will focus on the case of the black hole with the parameters in the range Eq. (\ref{range}).

In this spacetime, there are two Killing fields $\xi_{t, \phi}=\partial_{t, \phi}$ related to two conserved constants along the geodesics,
\begin{eqnarray}
 -\mathcal{E}&=&g_{\mu\nu}\xi_{t}^{\mu}p^{\nu},\label{e1}\\
 l&=&g_{\mu\nu}\xi_{\phi}^{\mu}p^{\nu},\label{e2}
\end{eqnarray}
where $\mathcal{E}$ and $l$ represent the energy and angular momentum of the particle, respectively, and $p^{\mu}=g_{\mu\nu}\dot{x}^{\nu}$ is the four-momentum of a test particle.

Next, we will employ the Lagrangian and Hamilton-Jacobi equation to obtain the equations of motion for a particle. The Lagrangian describing a neutral particle in the background (\ref{metric}) reads
\begin{eqnarray}
  \mathcal{L}&=&\frac{1}{2}g_{\mu\nu}\dot{x}^{\mu}\dot{x}^{\nu}\nonumber\\
             &=&\frac{1}{2}\bigg(-\frac{\Delta-a^{2}\sin^{2}\theta}{\rho^{2}}\dot{t}^{2}
                  -\frac{2a(r^{2}+a^{2}-\Delta)\sin^{2}\theta}{\rho^{2}}\dot{t}\dot{\phi}\nonumber\\
                 &&+\frac{\rho^{2}}{\Delta}\dot{r}^{2}
                  +\rho^{2}\dot{\theta}^{2}
                  +\frac{(r^{2}+a^{2})-\Delta a^{2}\sin^{2}\theta}{\rho^{2}}\sin^{2}\theta\dot{\phi}^{2}\bigg),\nonumber
\end{eqnarray}
where a dot over a symbol denotes the derivative with respect to an affine parameter $\lambda$. The normalizing condition is $g_{\mu\nu}\dot{x}^{\mu}\dot{x}^{\nu}=-\mu^{2}$ ($\mu^{2}$=-1, 0, and 1 for timelike, null, and spacelike geodesics, respectively). Regarding $x^{\mu}$ as the generalized coordinates, we then obtain the conjugate momenta
\begin{eqnarray}
  p_{\mu}=\frac{\partial \mathcal{L}}{\partial \dot{x}^{\mu}}=g_{\mu\nu}\dot{x}^{\nu}.
\end{eqnarray}
And the Hamiltonian reads
\begin{eqnarray}
  \mathcal{H}=p_{\mu}\dot{x}^{\mu}-\mathcal{L}
   =\frac{1}{2}g^{\mu\nu}p_{\mu}p_{\nu}.\label{Hamiltion}
\end{eqnarray}
Next, we will consider the Hamilton-Jacobi equation for geodesic motion in the background (\ref{metric}), which takes the
following form:
\begin{eqnarray}
 \frac{\partial \mathcal{S}}{\partial \lambda}=-\mathcal{H}=
         -\frac{1}{2}g^{\mu\nu}(\partial_{\mu}\mathcal{S})(\partial_{\nu}\mathcal{S}),
         \label{HJequation}
\end{eqnarray}
where $\mathcal{S}$ is the Hamilton-Jacobi function. With the form of the Hamilton-Jacobi function, one then can determine the generalized
coordinates $x^{\mu}$ as a function of time, which can uniquely determine the motion of a particle. To solve the Hamilton-Jacobi
equation, we first separate the Hamilton-Jacobi function as
\begin{eqnarray}
 \mathcal{S}=-\frac{\mu^{2}}{2}\lambda-\mathcal{E}t+l\phi+S_{r}(r)+S_{\theta}(\theta).
\end{eqnarray}
Substituting it into (\ref{HJequation}) and solving (\ref{e1}) and (\ref{e2}), we get
\begin{eqnarray}
 \rho^{2}\frac{d\phi}{d\lambda}
     &=&(l\csc^{2}\theta-a\mathcal{E})+\frac{a}{\Delta}\Big(\mathcal{E}(r^{2}+a^{2})-al\Big),\label{phiequation}\\
 \rho^{2}\frac{dt}{d\lambda}&=&a(l-a\mathcal{E}\sin^{2}\theta)
       +\frac{r^{2}+a^{2}}{\Delta}\Big(\mathcal{E}(r^{2}+a^{2})-al\Big),\\
 \rho^{2}\frac{dr}{d\lambda}&=&\sigma_{r}\sqrt{\Re},\label{Rad}\\
 \rho^{2}\frac{d\theta}{d\lambda}&=&\sigma_{\theta}\sqrt{\Theta},\label{thetaequation}
\end{eqnarray}
with ${\Re}$ and $\Theta$ given by
\begin{eqnarray}
 {\Re}&=&\left(a^2 \mathcal{E}-al+\mathcal{E} r^2\right)^2-\Delta
   \left(\mathcal{Q}+(l-a \mathcal{E})^2+r^2 \mu^2\right),\nonumber\\
 \Theta&=&\mathcal{Q}-a^2\mu^2 \cos^2\theta-(l\csc\theta-a
   \mathcal{E} \sin\theta)^2+(l-a \mathcal{E})^2.\nonumber
\end{eqnarray}
The sign functions $\sigma_{r}=\pm$ and $\sigma_{\theta}=\pm$ are independent from each other. Since this spacetime is of the Petrov type D, there are four conserved quantities along the geodesics of the test particle, $\mathcal{E}$, $l$, $\mu^{2}$, and $\mathcal{Q}$. In this paper, we adopt $\mu^{2}=0$ for discussing the black hole shadow.

\section{Effective potential and unstable circular orbits}
\label{potential}

In this section, we will give a brief study on the radial-motion of photons, which will be used to determine the contour of the noncommutative geometry inspired black hole shadow.

\begin{figure*}
\begin{center}
\subfigure[]{\label{pveff}
\includegraphics[width=8cm]{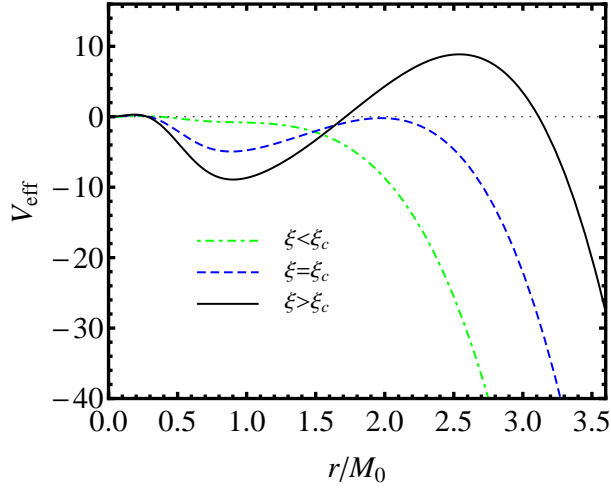}}
\subfigure[]{\label{pRc}
\includegraphics[width=8cm]{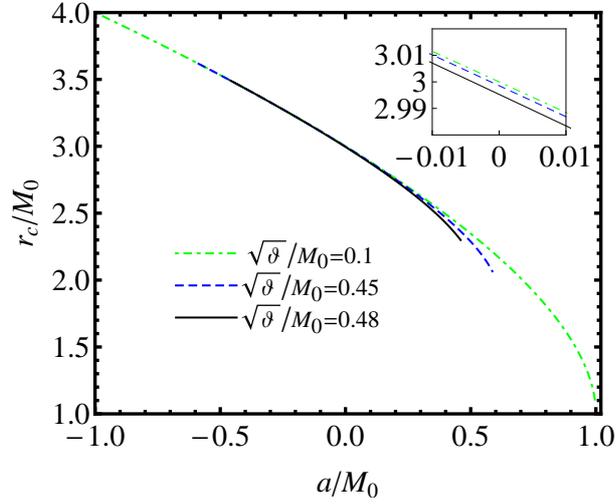}}
\end{center}
\caption{(a) Different behaviors of the effective potential $\mathcal{V}_{eff}$ for $\xi<\xi_{c}$, $\xi=\xi_{c}$, and $\xi>\xi_{c}$, respectively. (b) The radius $r_{c}$ of the equatorial circular orbit.}\label{pVR}
\end{figure*}

The radial motion (\ref{Rad}) can be written in the form
\begin{eqnarray}
 \bigg(\rho^{2}\frac{dr}{d\lambda}\bigg)^{2}+\mathcal{V}_{eff}(r)=0,
\end{eqnarray}
which is very similar to the equation of motion of a classical particle. The effective potential $\mathcal{V}_{eff}$ is given by
\begin{eqnarray}
 \mathcal{V}_{eff}(r)=&&\left(\eta+(\xi-a)^2\right)\bigg(r^{2}-4\frac{M_{0}r}{\sqrt{\pi}}\gamma(3/2, r^{2}/4\vartheta)+a^{2}\bigg)\nonumber\\
   &&-\left(a^{2}-a\xi+ r^2\right)^2,
\end{eqnarray}
where we have introduced two parameters $\eta=\mathcal{Q}/\mathcal{E}^{2}$ and
$\xi=l/\mathcal{E}$. The effective potential has the limit:
\begin{eqnarray}
\mathcal{V}_{eff}(0)=0 \quad \text{and} \quad
 { \mathcal{V}_{eff}(r\rightarrow\infty)\rightarrow-\infty}.
\end{eqnarray}
With a further analysis of $\mathcal{V}_{eff}(r)$, the motion of the test particle along the radial direction will be clear. We plot the effective potential $\mathcal{V}_{eff}$ against $r$ in Fig. \ref{pveff} with $\eta$, $\vartheta$, and $a/M_{0}$ fixed.

From the behavior of $\mathcal{V}_{eff}$ shown in Fig. \ref{pveff}, we get that there are three different kinds of orbit for a photon coming from infinity:
(i) $\xi>\xi_{c}$. The photon starting from infinity will meet a turning point, where $\mathcal{V}_{eff}(r)=0$, and then it turns back to infinity and be observed.
(ii) $\xi<\xi_{c}$. In this case, there is no turning point, thus the photon will cross the horizon and be absorbed by the black hole. Thus, the observer located at infinity will not observe such photon.
(iii) $\xi=\xi_{c}$. This is a critical case. The photon comes from infinity and approaches a turning point with zero radial velocity. The turning point corresponds to an unstable circular orbit. The radius of the orbit is determined by
\begin{eqnarray}
 \mathcal{V}_{eff}(r_{c})=\frac{\partial\mathcal{V}_{eff}}{\partial r}(r_{c})=0,\quad
 \frac{\partial^{2}\mathcal{V}_{eff}}{\partial r^{2}}(r_{c})<0.\label{vefff}
\end{eqnarray}
Here $r_{c}$ is an important parameter to determine the shape of the black hole. For the Schwarzschild black hole, $r_{c}=3M_{0}$; and for the Kerr black hole, $r_{c}=2M_{0}\big(1+\cos(\frac{2}{3}\arccos(\pm \frac{|a|}{M_{0}}))\big)$ for the equatorial circular orbit. For the noncommutative geometry inspired black hole, it cannot be written in an exact form, and we show it in Fig. \ref{pRc} for the equatorial circular orbit, from which we see that $r_{c}$ is almost the same for different values of $\vartheta$ with the spin parameter $a/M_{0}$ fixed. However, the deviation becomes notable when the black hole approaches an extremal one. With $r_{c}$ obtained, we can solve $\eta$ and $\xi$ form Eq. (\ref{vefff}):
\begin{eqnarray}
 \eta=\frac{A}{C^{2}},\quad \xi=\frac{B}{C},\label{ABC}
\end{eqnarray}
with the coefficients given by
\begin{eqnarray}
 A&=&-r_{c}^{10}-4 r_{c}^6 \vartheta ^{3/2} e^{\frac{r^2}{4 \vartheta }}
   \left[\sqrt{\pi } \left(2 a^2+r_{c}^2\right)-6 r_{c} \gamma
   \left(\frac{3}{2},\frac{r_{c}^2}{4 \vartheta
   }\right)\right]\nonumber\\
   &&-4 r_{c}^3 \vartheta ^3 e^{\frac{r_{c}^2}{2
   \vartheta }} \bigg[-4 \sqrt{\pi } \left(2 a^2+3
   r_{c}^2\right) \gamma \left(\frac{3}{2},\frac{r_{c}^2}{4
   \vartheta }\right)+\pi r_{c}^3\nonumber\\
   &&+36 r_{c} \gamma
   \left(\frac{3}{2},\frac{r_{c}^2}{4 \vartheta
   }\right)^2\bigg],\nonumber\\
 B&=&2 \vartheta ^{3/2} e^{\frac{r_{c}^2}{4 \vartheta }}
   \left[2
   \left(a^2-3 r_{c}^2\right) \gamma
   \left(\frac{3}{2},\frac{r_{c}^2}{4 \vartheta
   }\right)+\sqrt{\pi } r_{c} \left(a^2+r_{c}^2\right)\right]\nonumber\\
   &&+r_{c}^3
   \left(a^2+r_{c}^2\right),\nonumber\\
 C&=&a \left[r_{c}^3-2 \vartheta^{3/2} e^{\frac{r_{c}^2}{4 \vartheta }}
   \left(\sqrt{\pi} r_{c}-2 \gamma
   \left(\frac{3}{2},\frac{r_{c}^2}{4 \vartheta}\right)\right)\right].\nonumber
\end{eqnarray}
Using this expression, we will explore the black hole shadow in the next section by setting $M_{0}=1$ for simplicity.

\section{Black hole shadow}
\label{shape}

The boundary of the black hole shadow is determined by the unstable circular orbit. In the real observations, the shadow of a black hole is a dark region on the observer's sky. So it is natural to introduce the celestial coordinates
\begin{eqnarray}
 \alpha&=&\lim_{r\rightarrow \infty}
   \bigg(-r^{2}\sin\theta\frac{d\phi}{dr}
      \bigg|_{\theta\rightarrow i}\bigg)
     =-\xi\csc i,\nonumber\\
 \beta&=&\lim_{r\rightarrow \infty}
   \bigg(r^{2}\frac{d\theta}{dr}\bigg|_{\theta\rightarrow i}\bigg)
     =\pm\sqrt{\eta+a^{2}\cos^{2}i-\xi^{2}\cot^{2}i},\nonumber
\end{eqnarray}
where $i$ is the inclination angle between the axis of rotation of the black hole and the line of sight of the observer. The coordinates $\alpha$ and $\beta$ are, respectively, the apparent perpendicular distances of the image seen from the axis of symmetry and from its projection on the equatorial plane. After substituting into Eq. (\ref{ABC}), we get the explicit form of $\alpha$ and $\beta$:
\begin{eqnarray}
\alpha&=&-\frac{B}{C}\csc i,\\ \beta&=&\pm\frac{1}{C}\sqrt{A+a^{2}C^{2}\cos^{2}i+B^{2}\cot^{2}i}.
\end{eqnarray}

\subsection{Nonrotating case}

\begin{figure}
\begin{center}
\includegraphics[width=8cm]{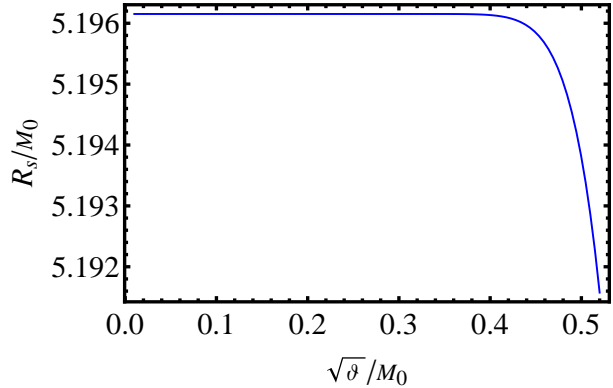}
\end{center}
\caption{The radius of the shadow for a nonrotating noncommutative geometry inspired black hole.}\label{prs}
\end{figure}

First, let us consider the nonrotating case, where the spin parameter $a/M_{0}=0$. A simple calculation shows
\begin{eqnarray}
 &&\alpha^{2}+\beta^{2}=\frac{2 r^2 \left(r^6+4 \vartheta ^3 e^{\frac{r^2}{2
   \vartheta }} \left(\pi  r^2-12 \gamma
   \left(\frac{3}{2},\frac{r^2}{4 \vartheta
   }\right)^2\right)-8 r^3 \vartheta ^{3/2}
   e^{\frac{r^2}{4 \vartheta }} \gamma
   \left(\frac{3}{2},\frac{r^2}{4 \vartheta
   }\right)\right)}{\left(r^3-2 \vartheta ^{3/2}
   e^{\frac{r^2}{4 \vartheta }} \left(\sqrt{\pi }
   r-2 \gamma \left(\frac{3}{2},\frac{r^2}{4
   \vartheta }\right)\right)\right)^2}. \label{ab1}
\end{eqnarray}
Since photons come from both sides of the nonrotating black hole have the same value of the deflection angle, the shadow of such case is a standard circle with radius $R_{s}/M_{0}=\sqrt{\alpha^{2}+\beta^{2}}$ given in Eq. (\ref{ab1}). It is easy to find that this radius depends of the inclination angle $i$, which is owing to the spherically symmetric of the black hole with vanishing spin. The radius of the shadow is also presented in Fig. \ref{prs}. For small value of $\sqrt{\vartheta}/M_{0}$, the radius almost equals to $R_{s}=3\sqrt{3}M_{0}$, which is exactly the result of the Schwarzschild black hole. Although the radius $R_{s}$ has a noticeable decrease at $\sqrt{\vartheta}/M_{0}=0.40$ from the figure, the difference is still indistinguishable from astronomical observations. For example, the difference between $\sqrt{\vartheta}/M_{0}=0.1$ and $0.48$ is 0.0013, between $\sqrt{\vartheta}/M_{0}=0.1$ and $0.52$ is 0.0046.

\subsection{Rotating case}

\begin{figure*}
\begin{center}
\subfigure[]{\label{pab1}
\includegraphics[width=6cm]{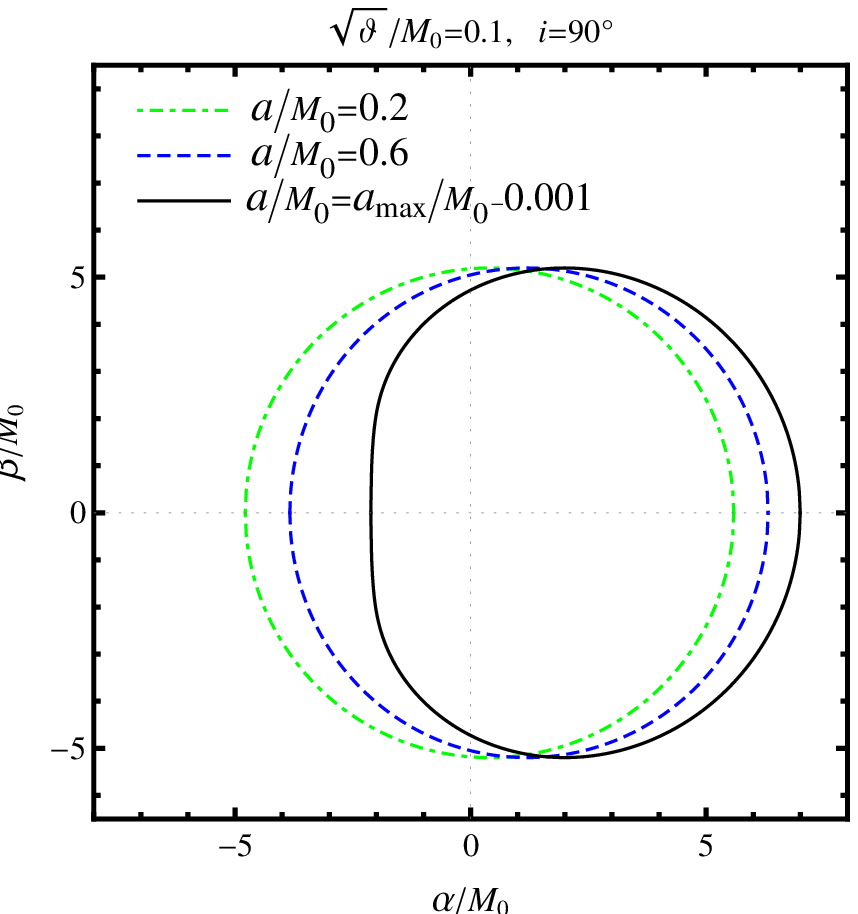}}
\subfigure[]{\label{pab2}
\includegraphics[width=6cm]{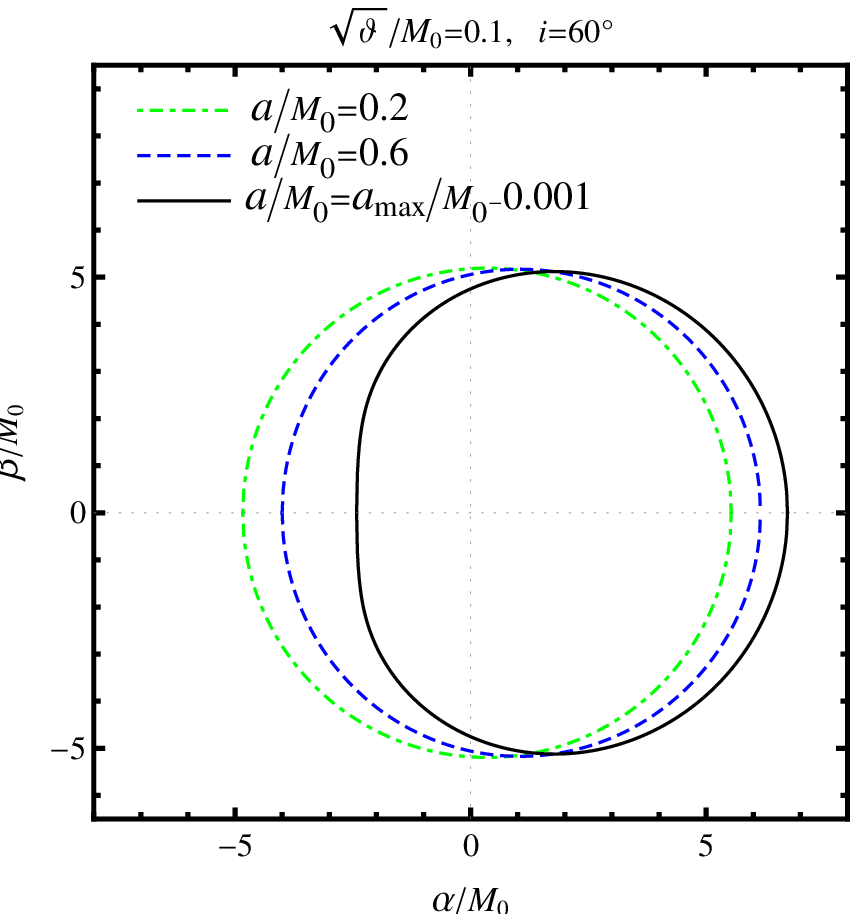}}\\
\subfigure[]{\label{pab3}
\includegraphics[width=6cm]{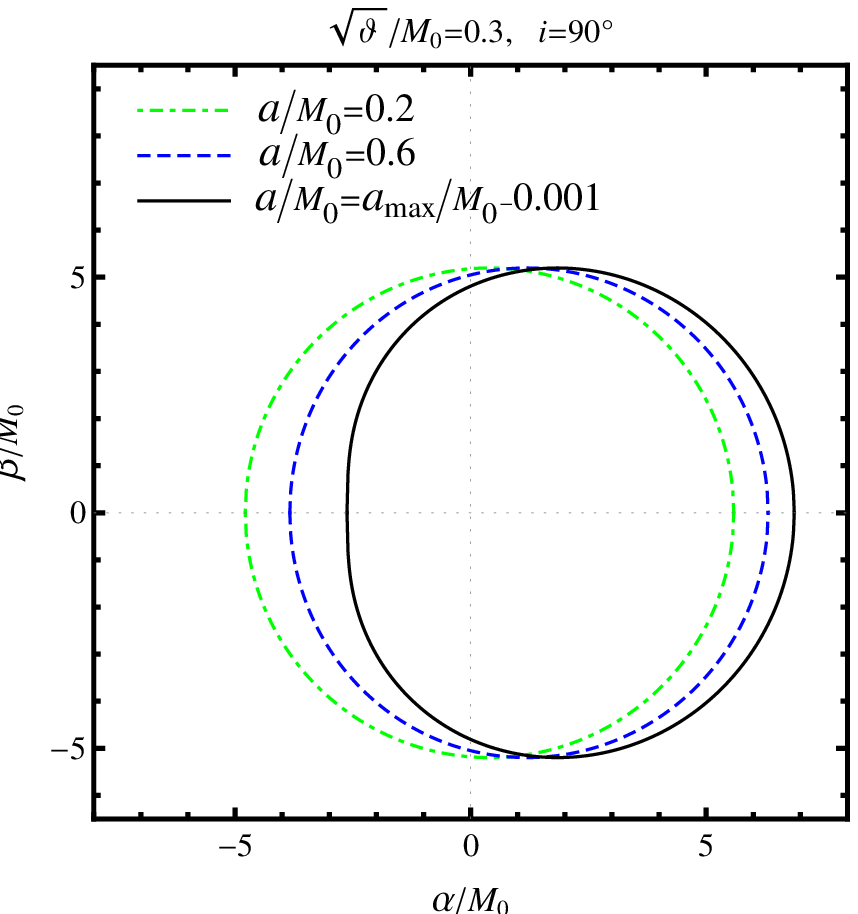}}
\subfigure[]{\label{pab4}
\includegraphics[width=6cm]{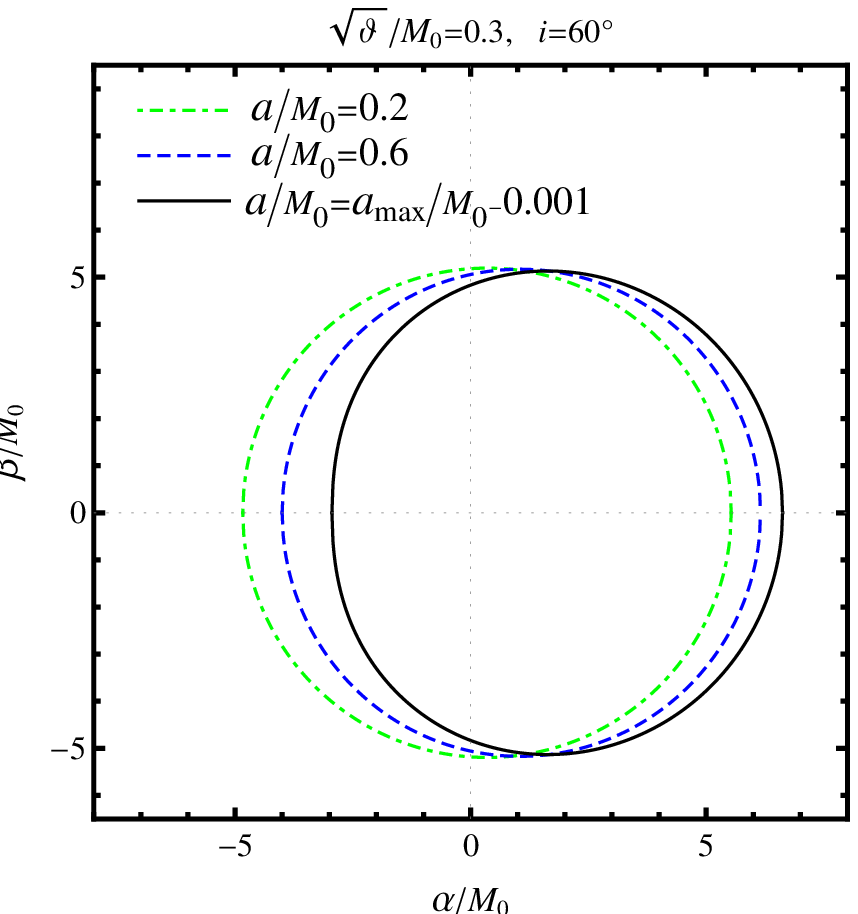}}\\
\subfigure[]{\label{pab5}
\includegraphics[width=6cm]{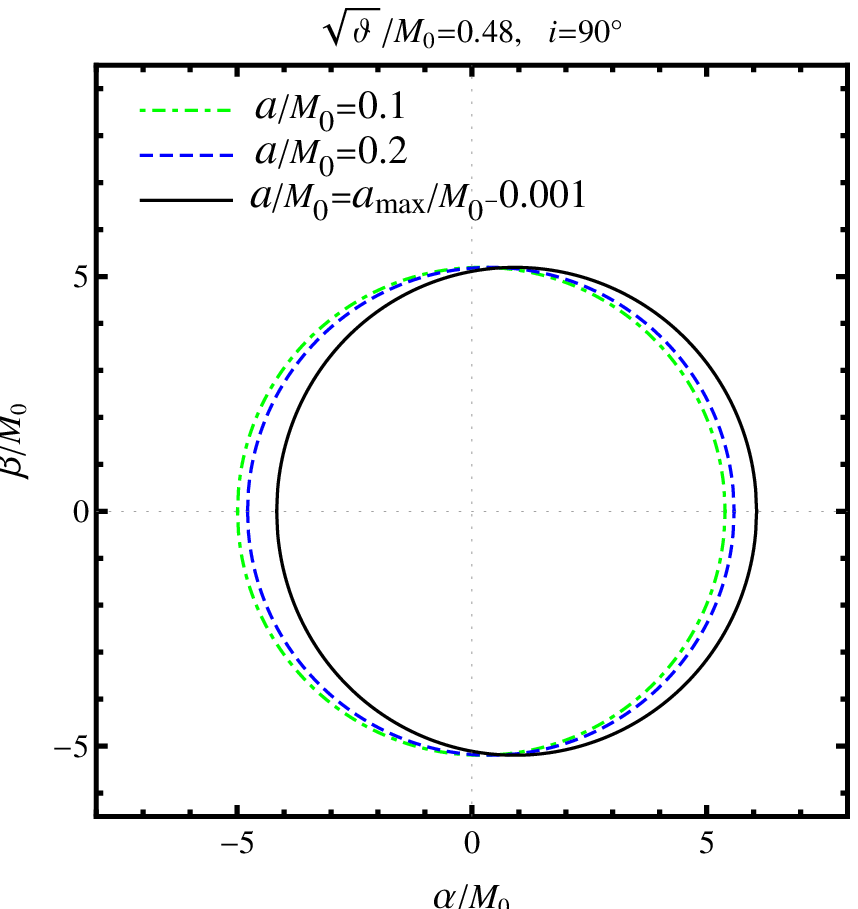}}
\subfigure[]{\label{pab6}
\includegraphics[width=6cm]{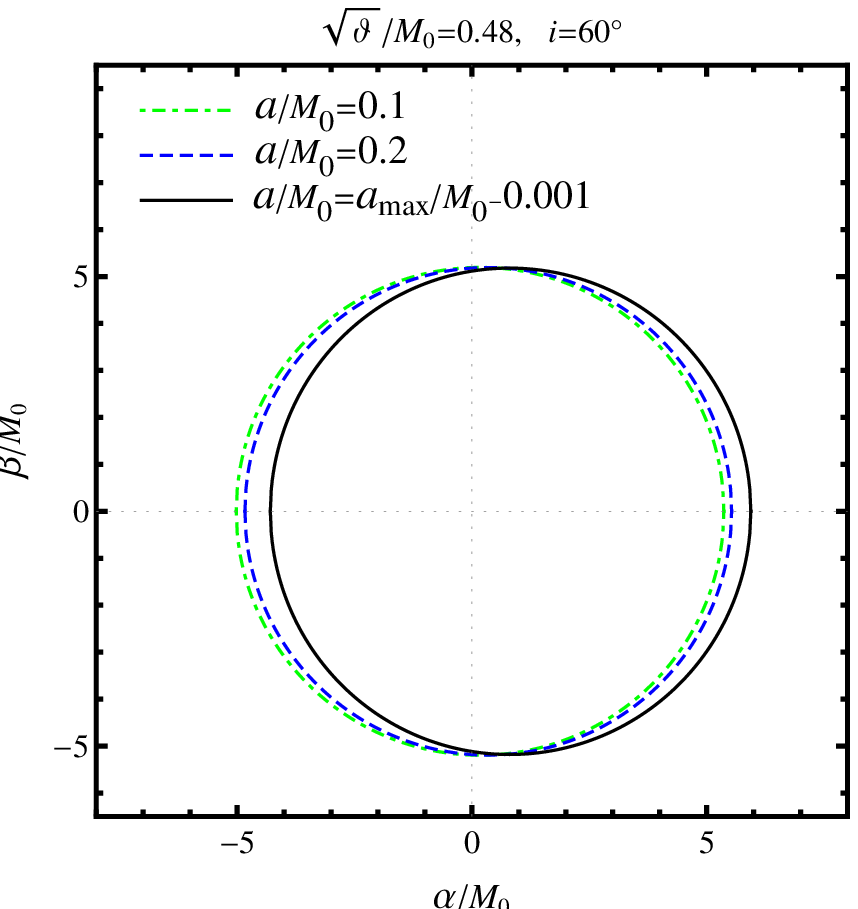}}
\caption{The shapes of the black hole shadow for different values of the parameters. The value of the parameter $a_{max}/M_{0}$ can be obtained by Eq. (\ref{thth}).}\label{pab}
\end{center}
\end{figure*}

For a rotating black hole, the photon capture radii for co-rotating and counter-rotating orbits will be different. From Fig. \ref{pRc}, we see that the radius of co-rotating orbits is always smaller than that of counter-rotating orbits. This is a universal result. For this reason, photons come from both sides of the black hole will have different values of the deflection angle, resulting to a deformed circle. The shape of the black hole shadow is presented in Fig. \ref{pab}. The shape depends on the parameter $\sqrt{\vartheta}/M_{0}$, $a/M_{0}$ and the inclination angle $i$. It is clear that for a slow rotating black hole, the shape is an approximate circle. While for a fast rotating black hole, the shape will deviate significantly from a circle. This deviation is found to increase with the inclination angle $i$. It is also an interesting result that the shape of the black hole shadow approaches a standard circle with the increase of $\sqrt{\vartheta}/M_{0}$, even for a near extremal black hole with small spin $a/M_{0}$.

In order to extract the information of an astronomical object from the shadow, the observables, which could be directly observed from the astronomical observation, are necessary. Let first focus on the silhouette of the shadow. Similar to the Kerr case, for each silhouette of the shadow, we clearly see that there are four characteristic points: the top point ($\alpha_{t}, \beta_{t}$), bottom point ($\alpha_{b}, \beta_{b}$), left point $(\alpha_{l}, 0)$, and right point ($\alpha_{r}, 0$) of the shadow, and they correspond to the unstable retrograde circular orbit seen from an observer located at infinity. Then a natural idea is that the observables can be constructed using these four points. In Ref. \cite{Maeda}, Hioki and Maeda proposed two observables, the radius $R_{s}$ and distortion $\delta_{s}$, to measure the shadow by a Kerr black hole. The observable $R_{s}$ is defined as the radius of a reference circle passing by the top, bottom, and right points. And $\delta_{s}$ gives its deformation with respect to the reference circle, measured by the left point $(\alpha_{l}, 0)$ and point $(\tilde{\alpha}_{r}, 0)$, which is the point where the reference circle cut the horizontal axis at the opposite side of $(\alpha_{r}, 0)$. Hence $R_{s}$ and $\delta_{s}$ measure the size and deformation of a shadow. With observables $R_{s}$ and $\delta_{s}$, the Kerr black hole spin $a/M_{0}$ and inclination angle $i$ of the observer can be uniquely determined. However for the noncommutative geometry inspired black hole, there exists an extra parameter $\sqrt{\vartheta}/M_{0}$. The presence of the noncommutative parameter will bring the degenerate problem as presented in Ref. \cite{Tsukamoto}, which means that different pairs of ($a/M_{0}$, $i$, $\sqrt{\vartheta}/M_{0}$) may have the same shadow. In order to solve this problem, Tsukamoto-Li-Bambi studied the fine structure of the shadow. They gave another distortion parameter $\varepsilon$ \cite{Tsukamoto}, which measures the distortion on the left side of the black hole shadow resulted by the unstable retrograde circular orbit. And one more point is introduced, which we denote as $(\alpha_{h}, \beta_{h})$. It is the point that the horizontal line of $\beta=\beta_{t}/2$ cuts the shadow at the opposite side of $(\alpha_{r}, 0)$. Combined with these three observables, the nature of black hole and inclination angle $i$ of the observer will be well determined for the noncommutative geometry inspired black hole.

\begin{figure*}
\begin{center}
\subfigure[]{\label{pdeltas1}
\includegraphics[width=6cm]{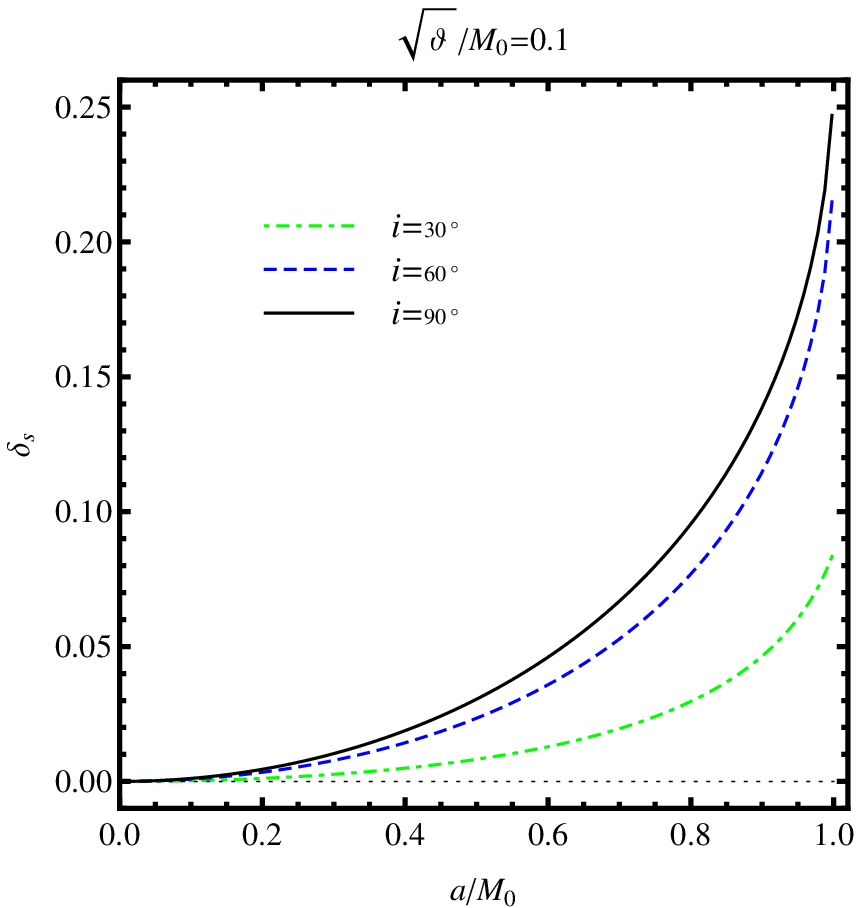}}
\subfigure[]{\label{pdeltas2}
\includegraphics[width=6cm]{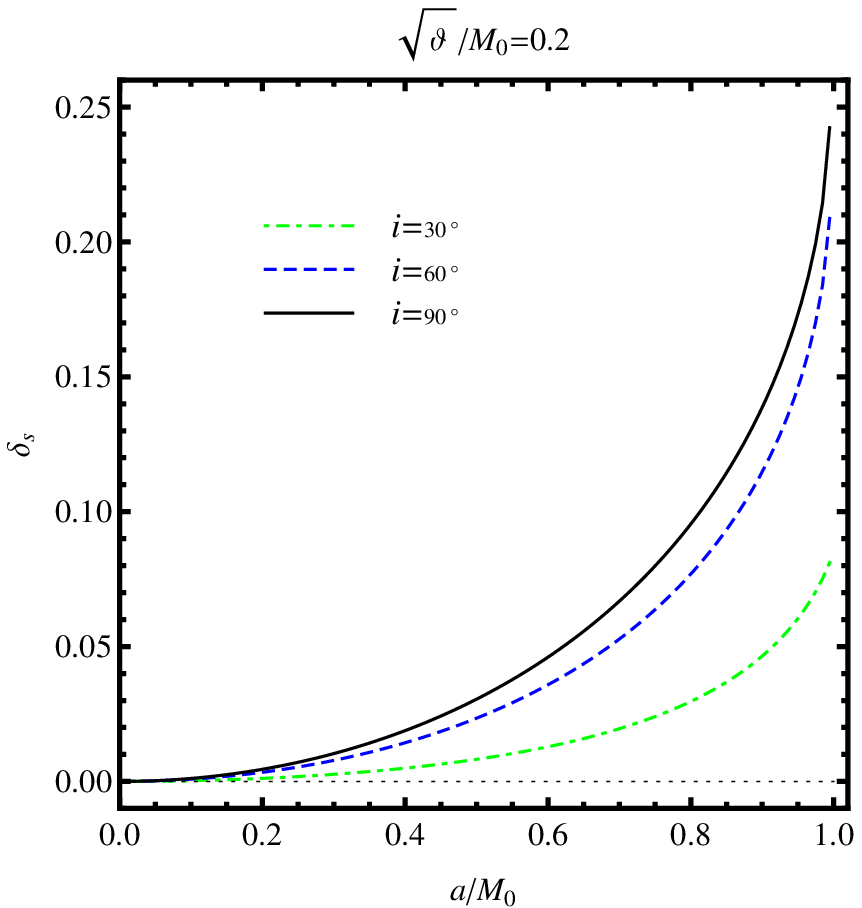}}\\
\subfigure[]{\label{pdeltas3}
\includegraphics[width=6cm]{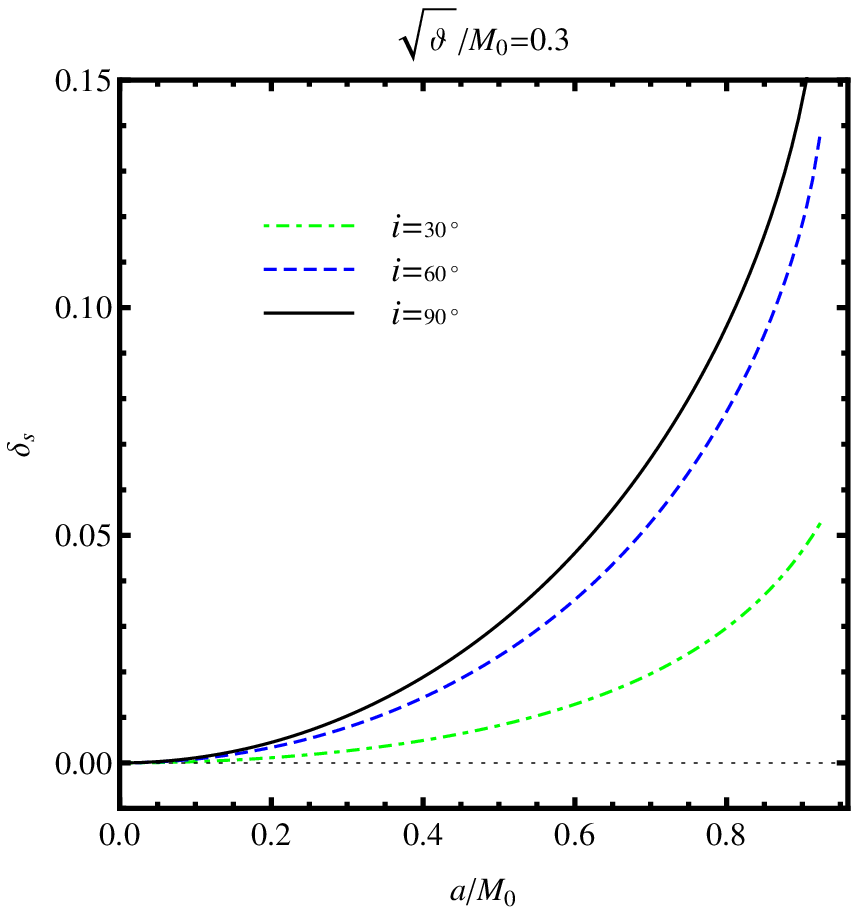}}
\subfigure[]{\label{pdeltas4}
\includegraphics[width=6cm]{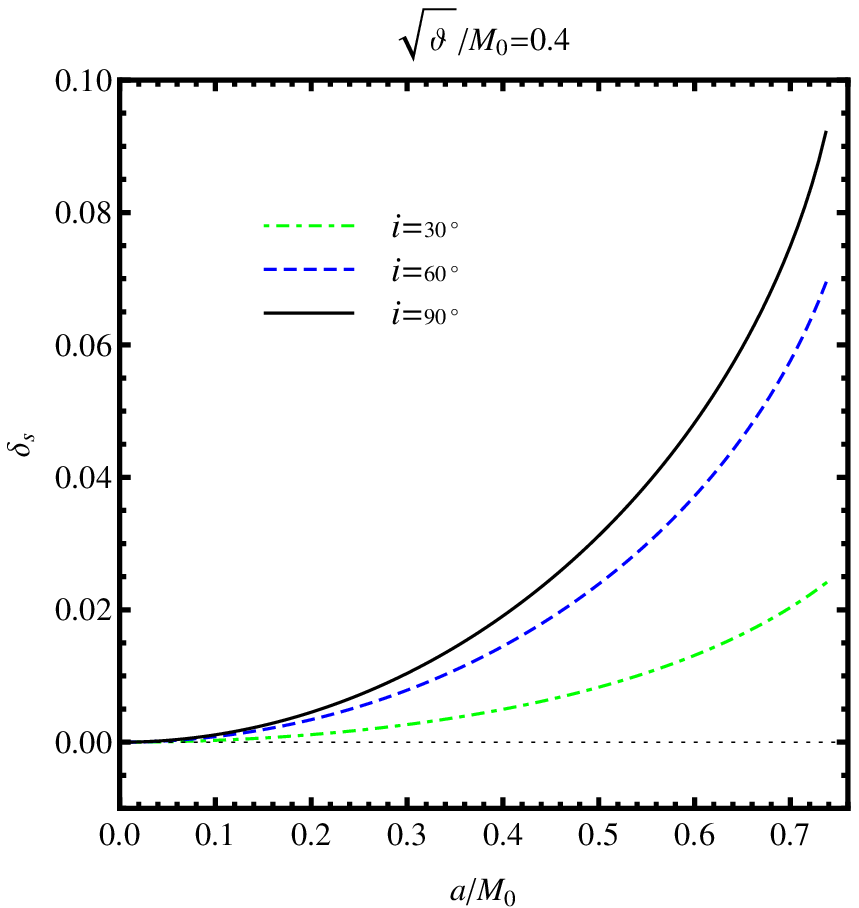}}\\
\subfigure[]{\label{pdeltas5}
\includegraphics[width=6cm]{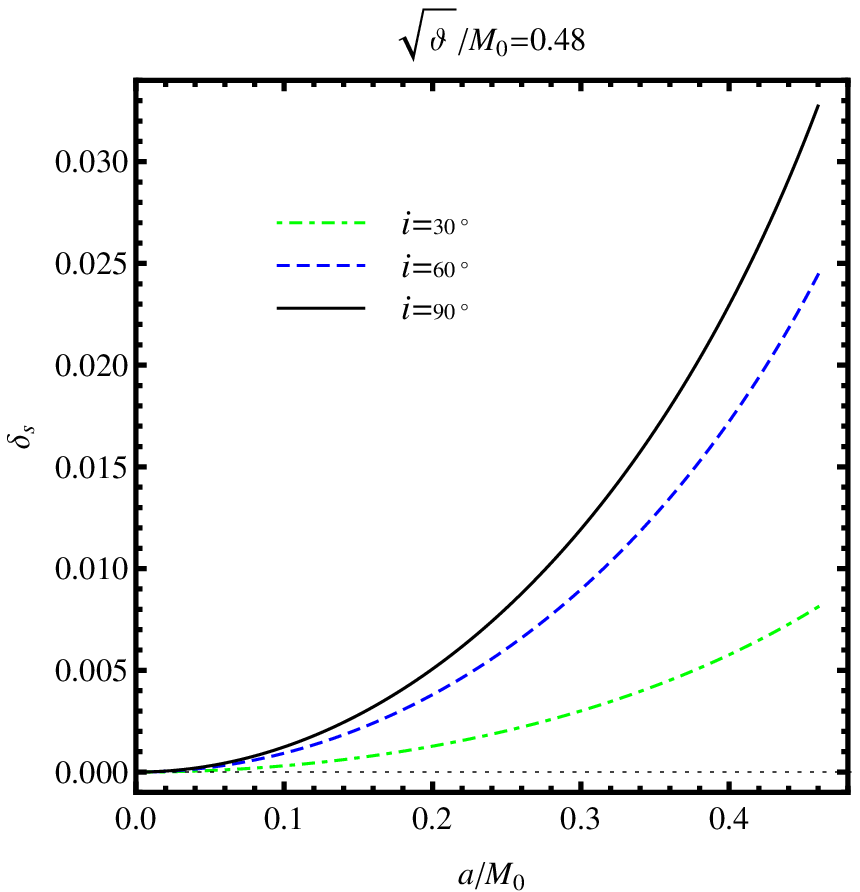}}
\subfigure[]{\label{pdeltas6}
\includegraphics[width=6cm]{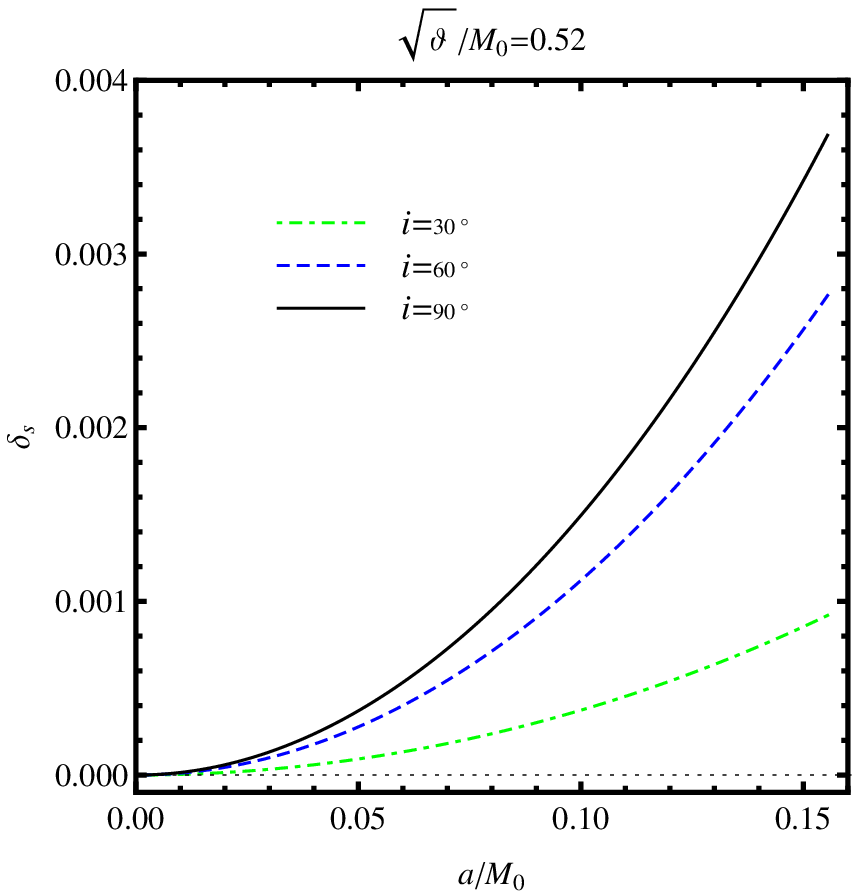}}
\caption{The distortion $\delta_{s}$ of the black hole shadow against spin $a/M_{0}$ for different values of the noncommutative parameter $\sqrt{\vartheta}/M_{0}$, and inclination angle $i$.}\label{ppdeltas}
\end{center}
\end{figure*}

\begin{figure*}
\begin{center}
\subfigure[]{\label{pe1}
\includegraphics[width=6cm]{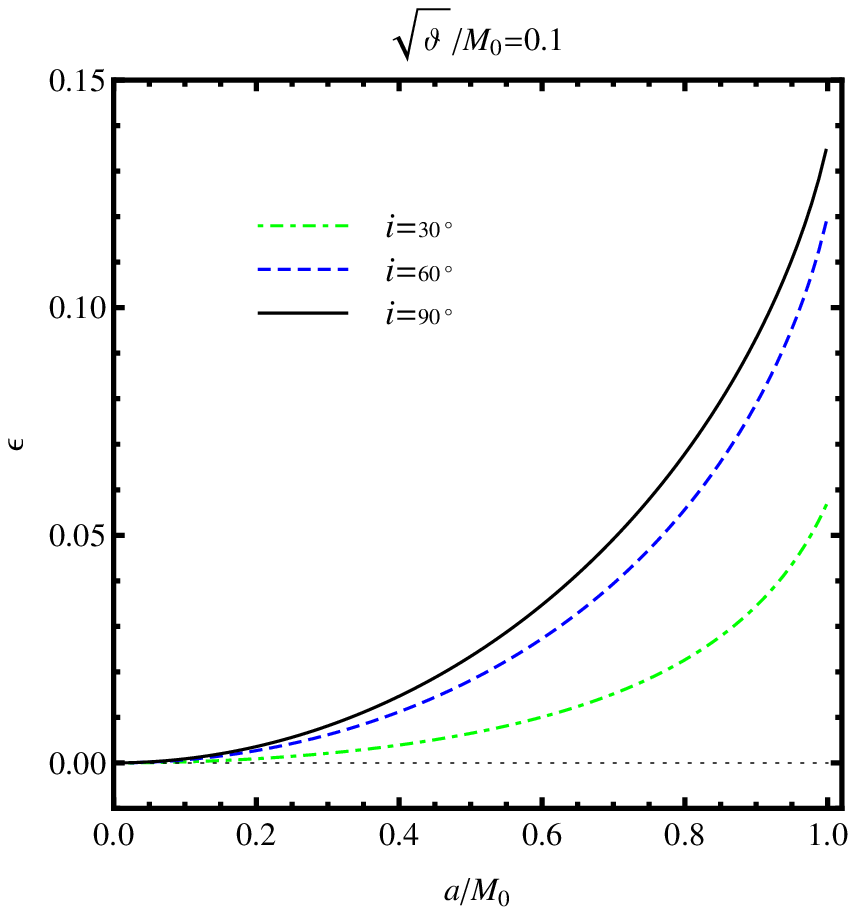}}
\subfigure[]{\label{pe2}
\includegraphics[width=6cm]{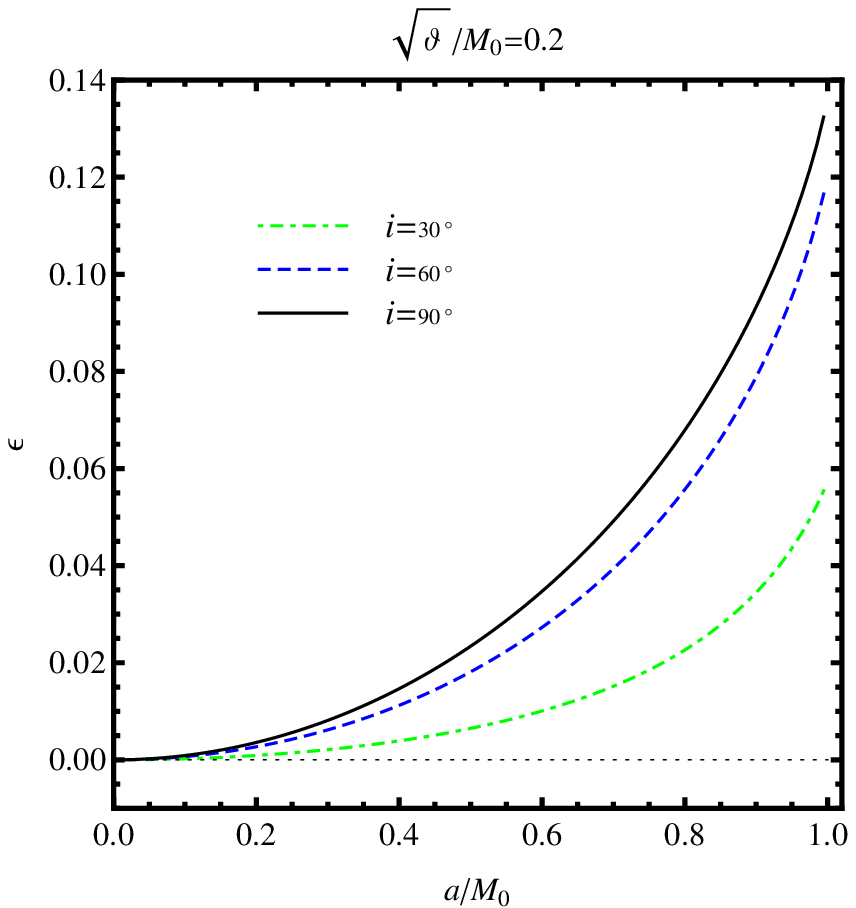}}\\
\subfigure[]{\label{pe3}
\includegraphics[width=6cm]{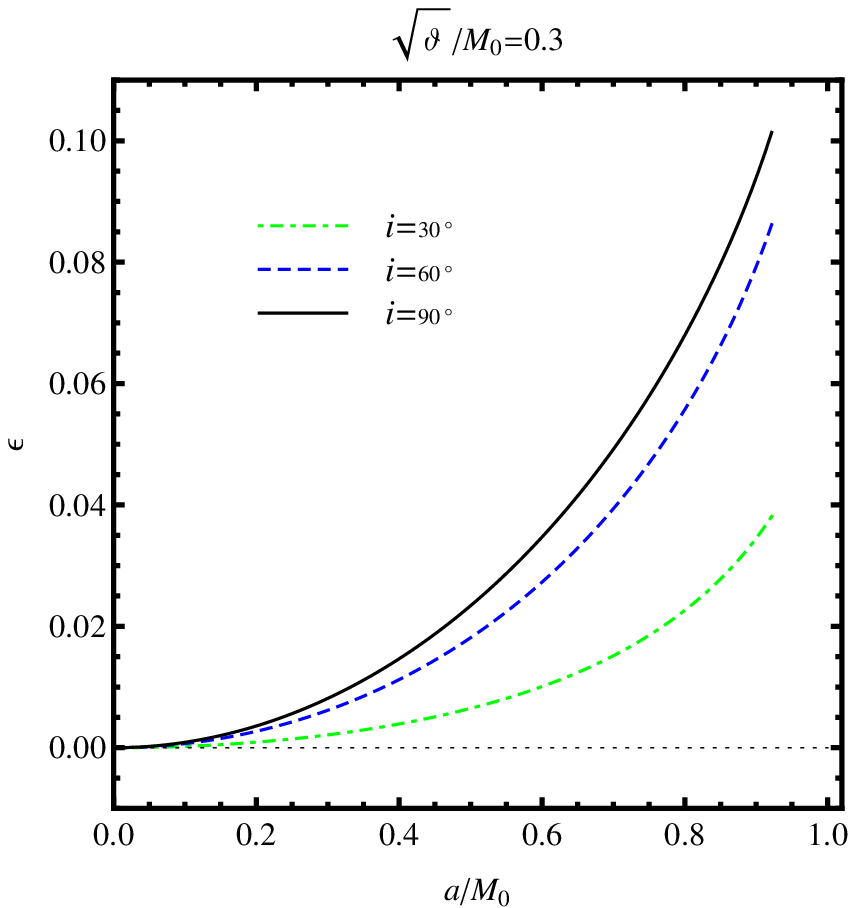}}
\subfigure[]{\label{pe4}
\includegraphics[width=6cm]{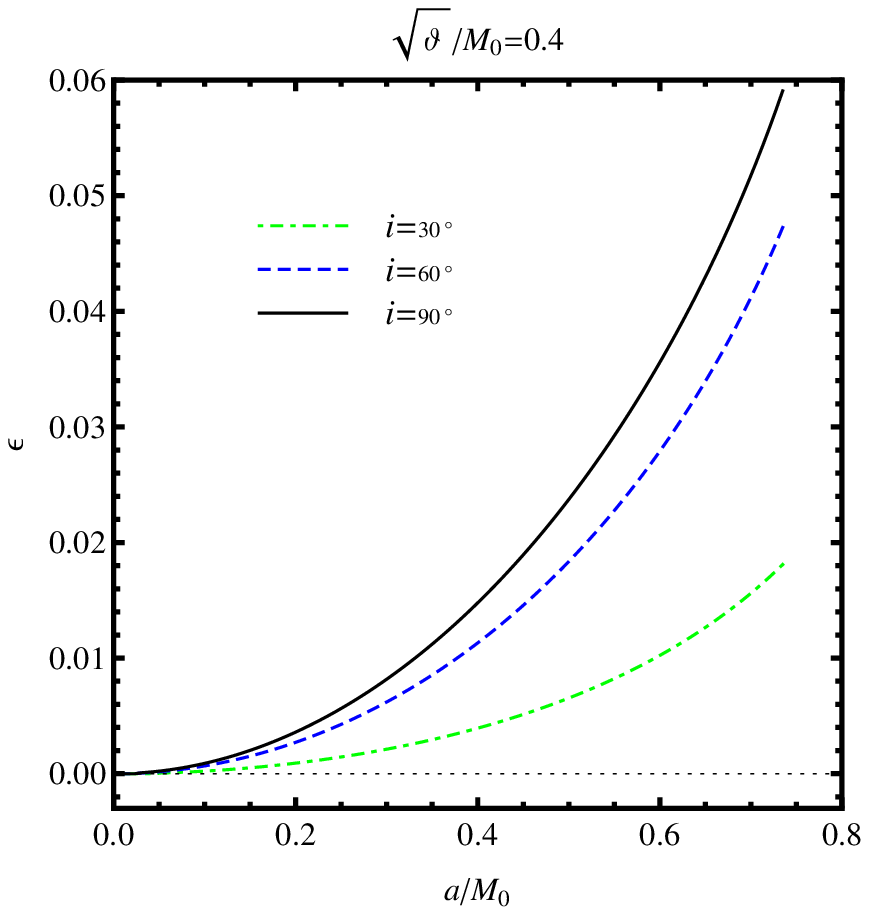}}\\
\subfigure[]{\label{pe5}
\includegraphics[width=6cm]{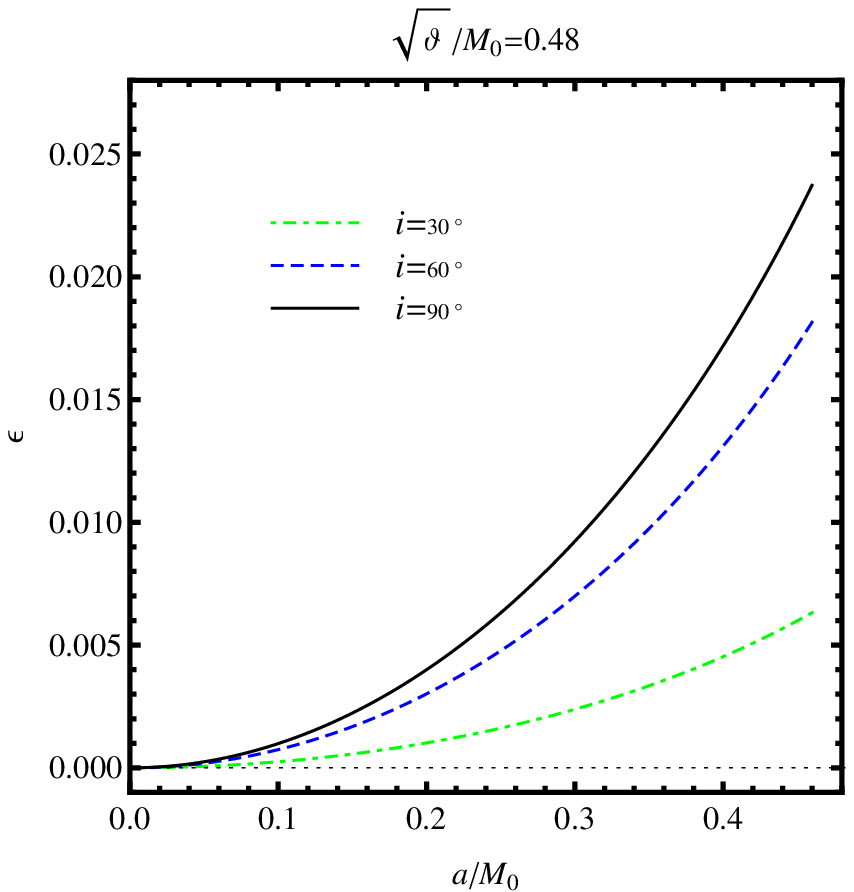}}
\subfigure[]{\label{pe6}
\includegraphics[width=6cm]{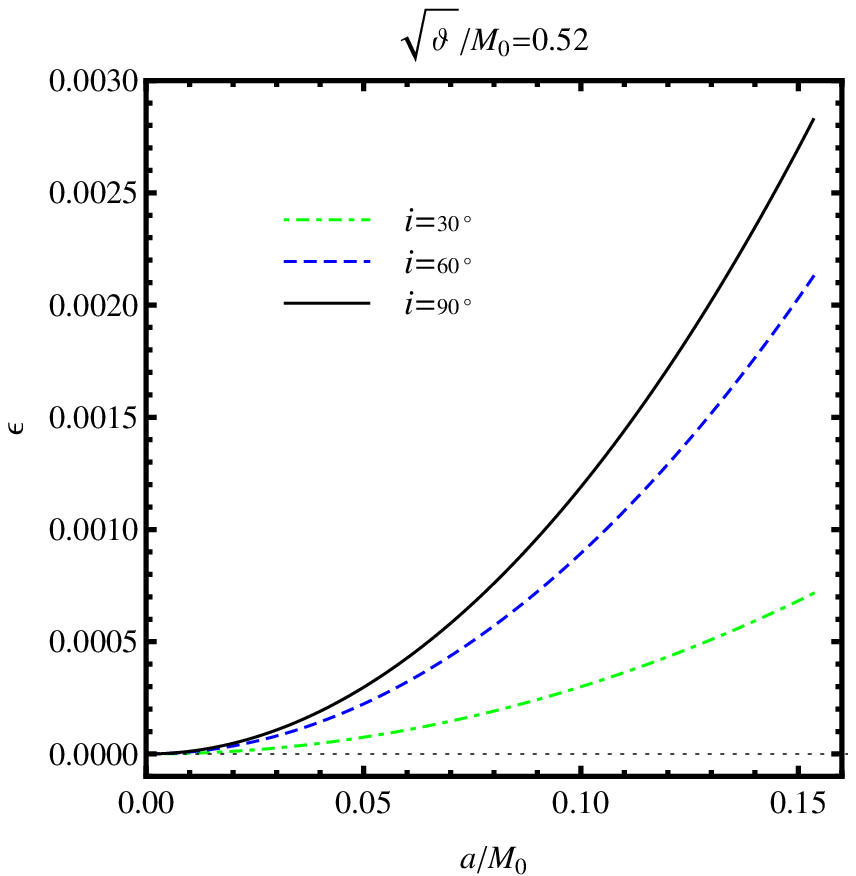}}
\caption{The distortion $\epsilon$ of the black hole shadow against spin $a/M_{0}$ for different values of the noncommutative parameter $\sqrt{\vartheta}/M_{0}$, and inclination angle $i$.}\label{ppe}
\end{center}
\end{figure*}

After a simple algebra calculation and combining with the symmetry of the shadow, we find that the center ($\alpha_{c}$, $\beta_{c}$) of the reference circle locates at
\begin{eqnarray}
 \alpha_{c}&=&\frac{\alpha_{r}^{2}-\alpha_{t}^{2}+\beta_{t}^{2}}{2(\alpha_{r}-\alpha_{t})},\\
 \beta_{c}&=&0.
\end{eqnarray}
Note that $\beta_{c}=0$ is caused by $\beta_{t}=\beta_{b}$. The expressions of the three observables read
\begin{eqnarray}
 R_{s}&=&\frac{(\alpha_{t}-\alpha_{r})^{2}+\beta_{t}^{2}}{2(\alpha_{r}-\alpha_{t})},
 \label{observable1}\\
 \delta_{s}&=&\frac{(\alpha_{l}-\tilde{\alpha}_{r})}{R_{s}},
 \label{observable2}\\
 \epsilon&=&1-\frac{(\alpha_{h}-\alpha_{c})^{2}+\beta_{t}^{2}/4}{R_{s}}.\label{observable3}
\end{eqnarray}
Here we have modified the distortion $\epsilon$ proposed by Tsukamoto-Li-Bambi \cite{Tsukamoto}. Then there will be no distortion when $\epsilon=0$. For the nonrotating noncommutative geometry inspired black hole, we find that $\sqrt{\vartheta}/M_{0}$ has a weak influence on $R_{s}$ (see Fig. \ref{prs}). This result also holds for the rotating black hole case. So it is very hard to distinguish a nonrotating noncommutative geometry inspired black hole from a Schwarzschild black hole, or distinguish a rotating noncommutative geometry inspired black hole from a Kerr black hole with the same spin through $R_{s}$. On the other hand, for a theoretical model, $R_{s}$ depends on the mass of the black hole and the distance from the observer. And these two quantities are known with a large uncertainty with modern astronomical observations, so the observable $R_{s}$ measuring the size of the shadow seems not to be a measurable quantity to probe the nature of a black hole. Compared with it, the distortions $\delta_{s}$ and $\epsilon$ measuring the shape of the shadow become a key point on testing the nature of the black hole. The behavior of $\delta_{s}$ is shown in Fig. \ref{ppdeltas} for different values of $a/M_{0}$, $\sqrt{\vartheta}/M_{0}$, and $i$. For the small value of $a/M_{0}$, i.e., $a/M_{0}<0.2$, $\delta_{s}$ approaches zero, which means the shape of the black hole shadow is almost a circle. For the large value of $a/M_{0}$, $\delta_{s}$ becomes large. And it takes the maximum value when the black hole approaches the extremal case. For example when $a/M_{0}=0.998$, $\delta_{s}\approx 25\%$ with $\sqrt{\vartheta}/M_{0}=0.1$ and $i=90^{\circ}$. Moreover, one can find that $\delta_{s}$ increases with the inclination angle $i$ when other parameters fixed. It is interesting to note that the maximum value of $\delta_{s}$ decreases with $\sqrt{\vartheta}/M_{0}$, which may be resulted by the decrease of the maximum spin $a_{max}$ with $\sqrt{\vartheta}/M_{0}$. However $\delta_{s}$ has a large value for large $\sqrt{\vartheta}/M_{0}$ with the spin $a/M_{0}$ fixed. We illustrate the distortion  $\epsilon$ in Fig. \ref{ppe}. It shares the same behavior as that of $\delta_{s}$. For fixed value of the black hole parameters, we find that $\epsilon$ is always smaller than $\delta_{s}$. If the inclination angle $i$ is given, then one of the distortions is enough to determine the nature of the black hole. Otherwise, the two distortions are useful to eliminate the degeneracy of $a/M_{0}$ and $\sqrt{\vartheta}/M_{0}$.

In order to compare it with that of the Kerr black hole, we list the numerical values of $\delta_{s}$ and $\epsilon$ in Tables \ref{tab1}-\ref{tab3} for $i=30^{\circ}$, $60^{\circ}$, and $90^{\circ}$, respectively. The result shows that for small value of $\sqrt{\vartheta}/M_{0}<0.2$, it is extremely hard to distinguish a noncommutative geometry inspired black hole from a Kerr black hole through the observation of the shadow shape. However, when $\sqrt{\vartheta}/M_{0}>0.4$, the difference becomes apparent. There may exist a deviation of $0.4\%\sim 1.0\%$ or even to several percentage. Therefore, the noncommutative parameter $\sqrt{\vartheta}/M_{0}\in (0.4, 0.52)$ is expected to be well-tested through the astronomical observations of the black hole shadow in the near future.

\small\begin{table}[h]
\begin{center}
\begin{tabular}{|c|c|c|c|c|c|c|}
  \hline
     & $a/M_{0}$=0.2 &  $a/M_{0}$=0.4 & $a/M_{0}$=0.5 & $a/M_{0}$=0.6 & $a/M_{0}$=0.7 \\\hline
  Kerr BH  & 0.1139/0.0914 &  0.4930/0.3916 & 0.8210/0.6481 & 1.286/1.007 & 1.956/1.516 \\
  $\sqrt{\vartheta}/M_{0}=0.20$& 0.1139/0.0914 &  0.4930/0.3916 & 0.8210/0.6481 & 1.286/1.007 & 1.956/1.516\\\hline
  $\sqrt{\vartheta}/M_{0}=0.40$& 0.1144/0.0917 &  0.4966/0.3942 & 0.8304/0.6544 & 1.312/1.023 & 2.035/1.561 \\\hline
  $\sqrt{\vartheta}/M_{0}=0.48$& 0.1272/0.1017 &  0.5765/0.4528 & .../... & .../... & .../... \\
  \hline
\end{tabular}
\caption{The distortions $\delta_{s}/\epsilon$ (\%)  of the shadows for the Kerr black hole and rotating noncommutative geometry inspired black hole with the inclination angle $i=30^{\circ}$.}\label{tab1}
\end{center}
\end{table}

\begin{table}[h]
\begin{center}
\begin{tabular}{|c|c|c|c|c|c|c|}
  \hline
     & $a/M_{0}$=0.2 &  $a/M_{0}$=0.4 & $a/M_{0}$=0.5 & $a/M_{0}$=0.6 & $a/M_{0}$=0.7 \\
\hline
  Kerr BH  & 0.3393/0.2704 &  1.435/1.123 & 2.345/1.813 & 3.584/2.729 & 5.280/3.939 \\\hline
  $\sqrt{\vartheta}/M_{0}=0.20$& 0.3393/0.2704 & 1.435/1.123 & 2.345/1.813 & 3.584/2.729 & 5.280/3.939\\\hline
  $\sqrt{\vartheta}/M_{0}=0.40$& 0.3408/0.2716 &  1.450/1.132 & 2.388/1.836 & 3.717/2.790 & 5.762/4.119 \\\hline
  $\sqrt{\vartheta}/M_{0}=0.48$& 0.3806/0.3013 &  1.724/1.311 & .../... & .../... & .../... \\
  \hline
\end{tabular}
\caption{The distortions $\delta_{s}/\epsilon$ (\%) of the shadows for the Kerr black hole and rotating noncommutative geometry inspired black hole with the inclination angle $i=60^{\circ}$.}\label{tab2}
\end{center}
\end{table}

\begin{table}[h]
\begin{center}
\begin{tabular}{|c|c|c|c|c|c|c|}
  \hline
     & $a/M_{0}$=0.2  & $a/M_{0}$=0.4 & $a/M_{0}$=0.5 & $a/M_{0}$=0.6 & $a/M_{0}$=0.7 \\
\hline
  Kerr BH  & 0.4507/0.3584  & 1.884/1.466 & 3.051/2.340 & 4.607/3.470 & 6.684 /4.917\\\hline
  $\sqrt{\vartheta}/M_{0}=0.20$& 0.4507/0.3584  & 1.884/1.466 & 3.051/2.340 & 4.607/3.470 & 6.684/4.917\\\hline
  $\sqrt{\vartheta}/M_{0}=0.40$& 0.4529/0.3599  & 1.907/1.478 & 3.119/2.372 & 4.822/3.558 & 7.496/5.174 \\\hline
  $\sqrt{\vartheta}/M_{0}=0.48$& 0.5067/0.3995  & 2.296/1.719 & .../... & .../... & .../... \\
  \hline
\end{tabular}
\caption{The distortions $\delta_{s}/\epsilon$ (\%) of the shadows for the Kerr black hole and rotating noncommutative geometry inspired black hole with the inclination angle $i=90^{\circ}$.}\label{tab3}
\end{center}
\end{table}

\section{Discussion}
\label{Discussion}

In this paper, we have investigated the shadow of the noncommutative geometry inspired black hole. The influence of the noncommutative parameter $\sqrt{\vartheta}/M_{0}$ on the shape of the black hole shadow was analyzed in detail. With the help of the null geodesics, the visualization of the black hole shadow was presented with the celestial coordinates $\alpha$ and $\beta$ for different values of the parameters.

For a nonrotating noncommutative geometry inspired black hole, the shape of the shadow is a perfect circle and its radius is found to depend on $\sqrt{\vartheta}/M_{0}$. For small value of $\sqrt{\vartheta}/M_{0}$, the radius $R_{s}$ almost equals to a constant $3\sqrt{3}M_{0}$, which is exactly the result of the Schwarzschild black hole. When $\sqrt{\vartheta}/M_{0}$ is larger than $0.4$, $R_{s}$ starts to decrease. However, even for $\sqrt{\vartheta}/M_{0}=0.52$, the radius $R_{s}$ deviates the constant $3\sqrt{3}M_{0}$ only of order $\sim 10^{-3}$, leading to a small variation in the size of the shadow. Thus, it is extremely hard to distinguish a nonrotating noncommutative geometry inspired black hole from a Schwarzschild black hole by measuring the size of the black hole shadow.

For a rotating noncommutative geometry inspired black hole, the shape of the shadow will be deformed due to the difference of the photon capture radius for the co-rotating and counter-rotating orbits. The change of the radius of the black hole shadow is very small for different values of the parameters. However the shadow will greatly deviate a circle when the black hole approaches to its extremal case of large spin $a/M_{0}$. For example, the distortion $\delta_{s}$ takes value $25\%$ for $a/M_{0}=0.998$. Thus, measuring the distortion $\delta_{s}$ of a shadow is an ideal test to determine the black hole spin $a/M_{0}$ and noncommutative parameter $\sqrt{\vartheta}/M_{0}$ of spacetime. The parameter $\delta_{s}$ is also found to increase with $\sqrt{\vartheta}/M_{0}$ with fixed $a/M_{0}$. After considering the fine structure of the shadow, we also studied the distortion $\epsilon$ first suggested in Ref. \cite{Tsukamoto}. We found it shares the similar behavior as that of $\delta_{s}$ while has a smaller value. Then with the three observables, $R_{s}$, $\delta_{s}$, and $\epsilon$, the quantities $a/M_{0}$ and $\sqrt{\vartheta}/M_{0}$, as well as the inclination angle $i$, could be uniquely determined.

Comparing with the Kerr black hole case, we saw that for small $\sqrt{\vartheta}/M_{0}$, it is extremely hard to distinguish the noncommutative geometry inspired black hole from the Kerr one. However, further increasing the noncommutative parameter such that $\sqrt{\vartheta}/M_{0}\in(0.4, 0.52)$, the distortion parameters $\delta_{s}$ and $\epsilon$ will have a deviations at the level $\sim 1\%$ or even to several percentage from the Kerr ones. And the difference may be measurable.

The observation of the black hole shadow is one of the main goals of the very long baseline interferometry. It is able to image the surrounding
environment of some supermassive Galactic black hole candidates, with resolution at the level of the black hole event horizon. And other observational facilities, such as the space-based RADIOASTRON and MAXIM, will also be able to observe the shadow. The nature of the supermassive Galactic black holes is expected to be determined through the astronomical observations in the near future. Meanwhile, considering the subtle effects of the noncommutative parameter $\sqrt{\vartheta}/M_{0}$ of spacetime discussed in this paper, the test of it may require a second generation of instruments with a high resolution.

\section*{Acknowledgements}

This work was supported by the National Natural Science Foundation of China (Grant No. 11205074 and Grant No. 11375075).

\end{document}